\documentclass[aps,pre,preprint,eqsecnum,showpacs]{revtex4-1}
\usepackage{amsmath}
\begin{document}

\title{Solitons and kinks in a general car-following model}
\date{June 21, 2012}

\author{ Douglas A. Kurtze }
\affiliation{ Department of Physics, Saint Joseph's University, 5600 City Avenue, Philadelphia, PA 19131, USA }

\begin{abstract}
We study a general car-following model of traffic flow on an infinitely long single-lane road, which assumes that a car's acceleration depends on its own speed, the headway between it and the car ahead, and the rate of change of headway, but makes minimal assumptions about the functional form of that dependence.  The velocity of uniform steady flow, as a function of headway, can be found implicitly from the acceleration function, and the linear stability criterion for that flow can be expressed simply in terms it.  Crucially, and unlike in previously analyzed car-following models, the threshold of absolute stability does {\it not\/} generally coincide with an inflection point in the steady-state velocity function.  The Burgers and Korteweg-deVries equations can be derived under the usual assumptions, but the modified Korteweg-deVries equation arises only when the threshold of absolute stability {\it does\/} coincide with an inflection point.  Otherwise, the Korteweg-deVries equation continues to apply near absolute stability, while near the inflection point one obtains the modified Korteweg-deVries equation plus an extra, quadratic term.  Corrections to the Korteweg-deVries equation ``select'' a single member of the a one-parameter set of one-soliton solutions by driving a slow evolution of the soliton parameter either away from or toward that member.  In previous models this selected soliton has always marked the threshold of a finite-amplitude instability of linearly stable steady flow, but here it can alternatively be a stable, small-amplitude jam that occurs when steady flow is linearly unstable.  The latter is the case when the Korteweg-deVries equation applies near absolute stability.  The new, augmented mKdV equation, which holds near an inflection point, admits a continuous family of kink solutions, like the modified Korteweg-deVries equation.  We derive the selection criterion arising from the corrections to this equation.
\end{abstract}


\maketitle

\section{\label{sec:intro}Introduction}

Car-following models constitute a class of highly idealized, ``microscopic'' models of vehicular traffic, which attempt to capture the characteristics of traffic flow by modeling the behavior of the individual cars.  A typical car-following model describes a single line of cars traveling along a long, uniform, straight or circular road.  Each driver is assumed to adjust his or her speed according to what the next car ahead is doing, and all drivers are taken to behave identically. It is well known that uniform, steady traffic flow can be unstable against the formation of slow-moving traffic jams; these models are used to identify conditions under which these jams form and to investigate their properties.

An important prototype of this class of models is the ``optimal velocity model'' (OVM) of Bando et al. \cite{Bando-PRE1995}, which is embodied in the equations
\[
  \tau\,{dv_n \over dt} = V_s(x_{n+1}-x_n) - v_n, \qquad v_n = dx_n/dt.
\]
Here the cars move in the positive $x$ direction, they are numbered consecutively, with car $n+1$ ahead of car $n$, and $x_n$ denotes the position of, say, the back of car $n$.  The ``optimal velocity'' function $V_s(h)$ is the speed at which a driver prefers to drive when the next car is ahead by a distance, or ``headway,'' of $h$.  The model assumes that each driver relaxes his or her car's speed to that preferred velocity with some fixed time constant $\tau$.  The optimal velocity function also acquires a second interpretation in this model, as it is trivial to show that $V_s(h)$ is the steady-state speed of uniform traffic.  That is, it is the speed at which every car moves when the cars are spaced an equal distance $h$ apart.

A model like the OVM, of course, ignores many features of the behavior of real human drivers, who not only have different reaction times, but also have individual preferred velocity functions which can change because of road conditions, fatigue, distractions, and the like.  However, such a model would give a better description of a line of cars controlled by on-board adaptive cruise control systems, which would use a small radar to monitor the headway to the next car.  The equation of motion would then be implemented by the programming of the cruise control.

The OVM was quickly generalized \cite{HelbingTilch-PRE1998} to account for the fact that the proper response to make to a car that is a distance $h$ ahead must also depend on that car's speed -- the response of car $n$ must be very different if the next car in line is stopped than it would be if that car is moving at some speed near $v_n$.  A particularly simple model that incorporates this effect is the ``full velocity difference model'' (FVDM) of Jiang et al. \cite{JiangWuZhu-PRE2001}, given by
\[
  \tau\,{dv_n \over dt} = V_s(x_{n+1}-x_n) - v_n + \lambda\,(v_{n+1}-v_n),
\]
where $\lambda$ is a constant which specifies the relative importance of the size of the headway and its rate of change in determining the response of car $n$.  Note that in this model, the function $V_s(h)$ retains its interpretation as the steady-state speed of uniformly spaced traffic, but it is {\it not\/} the speed to which an individual driver tends in all circumstances.  Instead, an individual driver tends to adjust his or her speed toward some weighted average of $V_s(h)$ and $v_{n+1}$, rather than just toward $V_s(h)$.  Tian, Jia, and Li \cite{TianJiaLi-CTP2011} suggested replacing the $v_{n+1}-v_n$ in the FVDM with a general function of $v_{n+1}-v_n$, naming their model the ``comprehensive optimal velocity model'' (COVM); they carried out calculations in the specific case where the function was a hyperbolic tangent.

A number of car-following models have been proposed in the literature; for reviews see \cite{Helbing-RMP2001,Nagatani-RPP2002,LiSun-JCTA2012}.  Analyses of different car-following models yield a range of results that are strikingly similar from model to model:

  (1) Steady, uniform flow with a fixed headway $h=\Delta$ is possible for a continuous range of headway values.  This flow is linearly stable or linearly unstable depending on whether $V_s'(\Delta)$, the derivative of the preferred velocity function, is below or above some critical value which depends on the parameters of the model.  (The preferred velocity function is often taken to have a sigmoidal, hyperbolic-tangent-like shape, in which case this means that light traffic, with a large spacing $\Delta$, is stable, and very heavy traffic, with small $\Delta$, is also stable albeit slow, while instability can happen for intermediate traffic densities. It is then possible to have {\it absolute stability}, when the slope $V_s'(\Delta)$ never exceeds the critical value.)  When $V_s'(\Delta)$ first crosses the critical value, the first disturbances to steady flow that become unstable are those with vanishingly small wave numbers.

  (2) If one assumes that the headway between cars varies only slightly from a uniform value $\Delta$, and that it varies slowly from car to car, then one can derive a Burgers equation for headway as a function of time and car number, as has been done for the OVM \cite{Nagatani-PRE2000}, the FVDM \cite{OuDaiDong-JPA2006}, and the COVM \cite{TianJiaLi-CTP2011}.  For those models, the coefficient of the nonlinear term in the Burgers equation is given by $V_s''(\Delta)$, and the diffusion coefficient is proportional to the difference between $V_s'(\Delta)$ and its critical value.  When uniform traffic is linearly stable, the diffusion coefficient is positive, and the Burgers equation then describes how slowly-varying deviations from uniformity diffuse away (via an interesting intermediate-asymptotic evolution \cite{BeckWayne-SIAMRev2011}).  When uniform flow is linearly {\it unstable}, the diffusion coefficient is negative, so the Burgers equation becomes a {\it backwards\/} diffusion equation.  Smooth initial perturbations then grow sharper, rapidly invalidating the assumptions under which the Burgers equation is derived.

  (3) When $V_s'(\Delta)$ is close to its critical value, one can reduce the model to a Korteweg-deVries (KdV) equation for the headway, plus small corrections; again, this has been done for the OVM \cite{MuramutsuNagatani-PRE1999}, the FVDM \cite{GeChengDai-PhysA2005,OuDaiDong-JPA2006}, and the COVM \cite{TianJiaLi-CTP2011}.  The KdV equation is completely integrable, and has soliton (and multi-soliton) solutions that can be written in closed form.  The correction terms, which break the integrability of the equation, have a very suggestive effect on the one-soliton solutions:  depending on the signs of certain coefficients in the correction terms, there may be at most one single member of the continuous family of one-soliton solutions that survives as a localized, shape-preserving solution.  Other members of the one-soliton family are slowly driven either toward or away from this one member \cite{KurtzeHong-PRE1995,ZhouLiuLuo-JPA2002}, again depending on signs of correction coefficients.  In general, however, there is a surviving soliton solution only when steady flow is linearly stable, and it marks the threshold of a finite-amplitude instability of that flow.

  (4) Near the onset of {\it absolute\/} stability -- that is, when there is at most a narrow range of parameters for which linear instability can occur -- the headway is described by a modified Korteweg-deVries (mKdV) equation, plus correction terms, in the OVM \cite{KomatsuSasa-PRE1995}, the FVDM \cite{GeChengDai-PhysA2005,OuDaiDong-JPA2006}, and the COVM \cite{TianJiaLi-CTP2011}.  This equation is also completely integrable; instead of localized soliton solutions, however, it has kink (and multi-kink) solutions which look like the transition zones between regions of different (linearly stable) traffic densities.  As is the case for one-soliton solutions of the KdV equation, there is a continuous family of one-kink solutions of the mKdV equation, and the correction terms select a single member of this family.

The purpose of this paper is to test the generality of these results.  For instance, the reduced equations (Burgers, KdV, mKdV) are derived from fairly long perturbation series about the steady-flow state.  Are the results of these derivations dependent on the fact that the OVM and FVDM assume that the acceleration of a car is a {\it linear\/} function of its velocity?  In the FVDM, the parameter $\lambda$, which specifies the relative importance of the size of the gap and its rate of change, is taken to be a constant.  Clearly it is unrealistic to expect the speed of car $n+1$ to be relevant if it is, say, several kilometers ahead of car $n$.  Do any important results change if it, or the time constant $\tau$, is allowed to depend on headway, as suggested by Gasser et al. \cite{GasserSeidelSiritoWerner-BIMAS2007}?  To answer questions such as these, we analyze a general possible car-following model, embodied in the equation \cite{Wilson-PTRSA2008}
\begin{equation}
  {dv_n \over dt} = A(x_{n+1}-x_n,v_{n+1}-v_n,v_n). \label{startpt}
\end{equation}
Here the acceleration function $A(h,\dot h,v)$ is a general function of the speed $v = v_n$ of the car under consideration, the headway $h = x_{n+1}-x_n$ between it and the next car ahead, and the rate $\dot h = v_{n+1}-v_n$ at which the headway is changing.  To be a realistic model of driver behavior, this function must satisfy some minimal conditions which we will discuss in Section \ref{sec:steady}.  Note that the model still assumes a single line of cars, so that the behavior of car $n$ depends only on the single car ahead of it, and it also assumes a uniform road, in that there is no position dependence in the acceleration function.  The model does {\it not\/} include explicit, finite time delays or look-ahead or look-back effects (viz., a dependence of the behavior of car $n$ on car $n+2$ or car $n-1$).

We find that many, but not all, of the results found in specific car-following models to date are in fact universal at the level of generality of (\ref{startpt}).  Specifically, the linear stability results and the derivations of the Burgers and KdV equations come out as usual.  However, a new term appears in the corrections to the KdV equation, and this term opens up the possibility of new behaviors not seen in its absence.  The same term shows up in the {\it leading\/} order in the derivation that usually produces the mKdV equation, so that the mKdV equation is {\it not\/} generic.  In the mKdV context, the possibility of having an extra term of this form was anticipated by Komatsu and Sasa \cite{KomatsuSasa-PRE1995}, and it was also found in a model including look-back by Hayakawa and Nakanishi \cite{HayakawaNakanishi-PRE1998}.

In Section \ref{sec:steady} we discuss the conditions that the function $A(h,\dot h,v)$ must satisfy in order for (\ref{startpt}) to be a reasonable model of traffic flow, and we carry out the linear stability analysis of steady traffic flow with a uniform headway $\Delta$.  We show how the analysis can be phrased in terms of the steady-state velocity function $V_s(\Delta)$, which, although it does not appear in the formulation of the model, can be defined implicitly from the acceleration function $A(h,\dot h,v)$.  We find that the linear stability results quoted above continue to hold at the level of generality of (\ref{startpt}), but that the condition for absolute stability must be stated in a more general way than usual.  In Section \ref{sec:longwave} we expand (\ref{startpt}) about the uniform steady state, assuming that the headway varies only slowly from car to car.  From the leading order of this expansion we find that the Burgers equation appears exactly as described above.  In Section \ref{sec:KdV} we modify the expansion from Section \ref{sec:longwave} to apply to a uniform steady state that is close to the onset of instability.  We find that the KdV equation appears at leading order as described above, but that an extra term appears in the coefficient of one of the corrections to the KdV equation.  We then carry out a multiple-time-scales analysis of the one-soliton solutions of the leading-order KdV equation.  We find that when the new term is absent, the unique one-soliton solution that is preserved by the corrections marks the onset of a finite-amplitude instability of a linearly stable uniform flow; with the new term present, however, it is possible instead for this one-soliton solution to represent a small-amplitude traffic jam that is stable when uniform flow is slightly unstable.  In Section \ref{sec:mKdV} we again modify the expansion from Section \ref{sec:longwave}, this time to apply to a uniform steady state that is close to an inflection point of the steady-state velocity function.  This normally corresponds to the onset of absolute stability, but at the level of generality of (\ref{startpt}) this need not be the case.  We find now that the new term appears at leading order -- that is, the flow is described by an evolution equation that is not the mKdV equation.  (The new term, however, does vanish if the inflection point coincides with the threshold of absolute stability.)  Like the mKdV equation, this new equation also has a one-parameter family of kink solutions.  Moreover, we show that it is still possible to carry out the selection analysis for these kink solutions, even though we have not been able to show that the new equation is integrable.  We discuss our results in Section \ref{sec:disc}.

\section{\label{sec:steady}Steady states and linear stability}

Before deriving and analyzing the steady states of the model embodied in (\ref{startpt}), we first set forth some conditions which a realistic acceleration function $A(h,\dot h,v)$ must satisfy \cite{Wilson-PTRSA2008}.  Here and throughout this paper we will assume that $A(h,\dot h,v)$ is differentiable as many times as necessary.  For a given headway $h$ and a given rate of change of headway $\dot h$, a driver will be more prone to decelerate or less prone to accelerate the faster his or her car is traveling.  Similarly, other things being equal a driver is more prone to accelerate the larger the headway is, or the more rapidly it is increasing.  Thus we expect to have
\begin{equation}
  {\partial A \over \partial v} \le 0, \qquad {\partial A \over \partial h} \ge 0,
  \qquad {\partial A \over \partial \dot h} \ge 0.
    \label{restrictions}
\end{equation}
We will in fact assume slightly more, namely that $\partial A/\partial v$ is strictly negative, or at least that there are no finite ranges of $v$ over which $A(h,0,v)$ is constant.

To find simple steady-state solutions of the model, we assume that the cars are equally spaced, and that they all cruise at the same speed, so they remain equally spaced.  If the headway in front of each car is $\Delta$, then this uniform flow is a steady state provided the common speed of the cars is $V_s(\Delta)$, which is defined implicitly by
\begin{equation}
  A(\Delta,0,V_s(\Delta)) = 0.
    \label{def:Vs}
\end{equation}
From the first inequality in (\ref{restrictions}) we see that $A(\Delta,0,v)$ is a nonincreasing function of $v$; given our assumption above that $A(h,0,v)$ does not remain constant over any finite range of $v$, the steady flow speed for a given $\Delta$ will then be unique.  Thus we have a one-parameter family of steady states of the model, labeled by the spacing $\Delta$; of course, these steady-flow solutions may be stable or unstable.

By differentiating (\ref{def:Vs}) repeatedly with respect to $\Delta$, we can obtain expressions for the derivatives of $V_s$.  For example, we find
\begin{equation}
  V_s'(\Delta) = -{ A_h(\Delta,0,V_s(\Delta)) \over A_v(\Delta,0,V_s(\Delta)) },
    \label{Vsprime}
\end{equation}
where, here and henceforth, primes denote derivatives with respect to $\Delta$ and subscripts denote partial derivatives, e.g. $A_h \equiv \partial A/\partial h$.  Given the restrictions (\ref{restrictions}) on the signs of the derivatives of $A$, we see that $V_s'(\Delta)$ cannot be negative.  By differentiating (\ref{Vsprime}), we find
\begin{equation}
  V_s''(\Delta) \equiv { d^2V_s \over d\Delta^2 } = -{A_{hh} + 2 A_{hv} V_s'(\Delta) + A_{vv} V_s'^2(\Delta) \over A_v},
    \label{Vs2prime}
\end{equation}
in which (again, here and henceforth) is it to be understood that the partial derivatives are evaluated in the steady state $(h = \Delta, \dot h = 0, v = V_s(\Delta))$.  In like manner we can also derive expressions for higher derivatives of $V_s$.

The linear stability of these steady-flow solutions has been analyzed at this level of generality by Wilson \cite{Wilson-PTRSA2008} and by Orosz et al. \cite{OroszWilsonStepan-PTRSA2010}; we recapitulate that calculation here, mainly to fix notation and to make contact with the linear stability results for specific models in the literature.  We proceed in the usual way, first writing
\begin{equation}
  x_n = n\,\Delta + V_s(\Delta)\,t + \hat x_n(t),
\end{equation}
where $\hat x_n$ represents an infinitesimal deviation of the car positions from the steady-flow solution.  We substitute this into the equation of motion (\ref{startpt}) and linearize in the $\hat x_n$ to get
\begin{equation}
  { d^2\hat x_n \over dt^2 } = {V_s'(\Delta) \over \tau} \, (\hat x_{n+1} - \hat x_n) -
  { 1 \over \tau } \, { d\hat x_n \over dt } + { \lambda \over \tau } \, { d \over dt } (\hat x_{n+1} - \hat x_n),
    \label{linearization}
\end{equation}
where the new parameters $\tau$ and $\lambda$ are defined by
\begin{equation}
  { 1 \over \tau } = -A_v(\Delta,0,V_s(\Delta)), \qquad
  { \lambda \over \tau } = A_{\dot h}(\Delta,0,V_s(\Delta)).
    \label{def:lambda,tau}
\end{equation}
These definitions agree with the notation of Ou et al. \cite{OuDaiDong-JPA2006}, but unlike in that paper $\lambda$ and $\tau$ can now be nontrivial functions of the uniform spacing $\Delta$.  We see from the inequalities (\ref{restrictions}) that they must be positive (or at least non-negative).  Note for future reference that (\ref{Vs2prime}), with this notation, reduces to
\begin{equation}
  V_s''(\Delta) = \tau(\Delta) [A_{hh} + 2 A_{hv} V_s'(\Delta) + A_{vv} V_s'^2(\Delta)].
    \label{betterVs2prime}
\end{equation}
Similarly, we find
\begin{equation}
  \left({ 1 \over \tau }\right)' = -[A_{hv} + A_{vv} V_s'(\Delta)], \qquad
  \left({ \lambda \over \tau }\right)' = [A_{h\dot h} + A_{\dot h v} V_s'(\Delta)],
\end{equation}
and so forth.

Since the coefficients in the linearized evolution equation (\ref{linearization}) are independent of car index $n$ and time $t$, we write the deviation $\hat x_n(t)$ as a linear combination of Fourier modes $\exp(ikn)$, with amplitudes that grow or decay exponentially in time, as $\exp[\sigma(k)t]$.  We then find that the (complex) linear growth rates $\sigma(k)$ are given by
\begin{equation}
  \sigma^2 \tau(\Delta) = V_s'(\Delta) (e^{ik} - 1) - \sigma + \lambda(\Delta) \sigma (e^{ik} - 1).
   \label{dispersionrel}
\end{equation}
A mode of wave number $k$ is marginally unstable when its growth rate $\sigma(k)$ is pure imaginary.  Separating the real and imaginary parts of (\ref{dispersionrel}) then shows, after some algebra, that this occurs for
\begin{equation}
  V_s' = { 1 + 2 \lambda \over 2 \tau } \, \left[ 1 + (1 + 2 \lambda) \tan^2 {k\over2} \right],
   \label{kismarginal}
\end{equation}
and that when the parameters satisfy this condition, the two solutions of the quadratic (\ref{dispersionrel}) are
\begin{equation}
  \sigma\tau = i (1+2\lambda) \tan{k\over2} \qquad \text{and} \qquad
    \sigma\tau = - \left( 1 + 2 \lambda \sin^2{k\over2} \right) \, \left( 1 + i \tan{k\over2} \right).
\end{equation}
Clearly, the second solution has negative real part, and so represents a stable mode; the first can easily be shown to acquire a positive real part when $V_s'$ is {\it larger\/} than the critical value given in (\ref{kismarginal}).  The steady state is unstable if {\it any\/} perturbation grows, so it is unstable if (for a fixed $\Delta$) $V_s'$ exceeds the {\it smallest\/} possible value of the right side of (\ref{kismarginal}), which is reached at $k=0$.  Thus we find that the steady state is
\begin{equation}
  \text{linearly unstable for} \qquad V_s'(\Delta) > \Omega_c(\Delta) \equiv { 1 + 2 \lambda(\Delta) \over 2 \tau(\Delta) },
   \label{stabilitycriterion}
\end{equation}
and if $V_s'$ is increased past the instability limit $\Omega_c$, then it is the long-wavelength ($k=0$) perturbations that first become unstable.

Depending on the details of the acceleration function, it is possible for the steady-flow states to be ``absolutely stable,'' in the sense that there is no $\Delta$ for which the flow is linearly unstable.  This is the case if $V_s'(\Delta)$ is always less than $\Omega_c(\Delta)$.  The {\it threshold\/} of absolute stability would be where only a single $\Delta$ is marginally stable, i.e. where $V_s - \Omega_c$ has a maximum at a value of zero.  For this to happen, there must be a $\Delta$ for which the value and slope of $V_s'$ match the value and slope of $\Omega_c$, and for which $V_s'''$ is less than (or more negative than) the value of $\Omega_c''$.  In the OVM, the FVDM, and the COVM the parameters $\tau$ and $\lambda$ are constants independent of $\Delta$, and so $\Omega_c$ is a constant also.  Thus in those models the threshold of absolute stability occurs at an inflection point of $V_s$ -- but in more general models this need not be the case.

Whether or not $V_s'$ is near the threshold of instability, it is easy to extract the behavior of long-wavelength perturbations from (\ref{dispersionrel}).  We immediately see that for small $k$, one of the solutions has $\sigma \approx -1/\tau$, so that this represents a decaying perturbation.  The other solution can be found by expanding in powers of $k$, and is given by
\begin{eqnarray}
  \Im \sigma &=& V_s' k + O(k^3), \nonumber \\
  \Re \sigma &=& \left[ \tau V_s' - \lambda - {1\over2} \right] V_s' k^2 + O(k^4).
   \label{sigmaatonset}
\end{eqnarray}
From the real part, we see that this perturbation grows or decays according to whether $V_s'$ is above or below the threshold given in (\ref{stabilitycriterion}).  (When it is {\it at\/} the threshold, the neglected $k^4$ term turns out to be negative, so the perturbation then decays.)  From the imaginary part, we see that the perturbation moves as it grows or decays, at a rate (in {\it cars\/} per unit time) of $V_s'(\Delta)$ in the upstream direction relative to the cars.  Since the cars are a distance $\Delta$ apart, at least before the perturbation develops, this is a {\it distance\/} per unit time of $\Delta V_s'(\Delta)$ relative to the cars.  And since the cars are moving in the downstream direction at $V_s(\Delta)$ relative to the road, the perturbation moves relative to the road at a velocity of approximately $V_s(\Delta) - \Delta V_s'(\Delta)$, which may be positive (downstream) or negative (upstream).

The stability criterion (\ref{stabilitycriterion}) is identical to that obtained by Ou et al. \cite{OuDaiDong-JPA2006} for the FVDM, except for the more general definitions of $V_s'$, $\tau$, and $\lambda$ given in (\ref{Vsprime}) and (\ref{def:lambda,tau}).  This should come as no surprise, since their model is essentially the general model expanded to linear order in the car velocities.  Thus linear stability analysis is not sensitive to the generalizations implied in our evolution equation (\ref{startpt}).

\section{\label{sec:longwave}Long-wave perturbations}

As we saw from the linear stability analysis, when traffic conditions are such that steady flow is just on the verge of instability, it is the long-wavelength perturbations that are the first to grow.  It is therefore of interest to approximate the full model in a way that focuses on these perturbations.  To this end, we assume that the deviation of the cars' actual positions from their steady-flow positions varies slowly from one car to the next.  From the small-$k$ linear stability result (\ref{sigmaatonset}) we see that each mode with a small wave number $k$ grows or decays at a rate proportional to $k^2$, while moving relative to traffic at a rate -- in units of cars per unit time -- of $-V_s'(\Delta)$.  Motivated by these considerations we write
\begin{equation}
  x_n = n\Delta + V_s(\Delta)t + f(z,T),
   \label{ansatz}
\end{equation}
where
\begin{equation}
  z \equiv \epsilon [n + V_s'(\Delta) t], \qquad T \equiv \epsilon^2 t.
   \label{def:z,T}
\end{equation}
Here, $\epsilon$ is a small but otherwise arbitrary parameter and $f$ and its derivatives are taken to be of order unity.  Thus in order for $f$ to {\it change\/} by an amount of order unity, the car number $n$ must change by an amount of order $\epsilon^{-1}$, so that $\epsilon$ measures the slowness of the variation of the position deviation $f$ along the line of cars.  This ansatz amounts to assuming that there is a pattern of deviations of headway from the uniform value $\Delta$, that this pattern propagates upstream through the line of cars at a rate $V_s'(\Delta)$, and that its shape changes slowly as it propagates.

If $f$ is of order unity, then the velocities of the cars and the spacings between them deviate from their steady-flow values by order $\epsilon$,
\begin{eqnarray}
  v_n &=& V_s(\Delta) + \epsilon V_s'(\Delta) f_z + \epsilon^2 f_T, \\
  x_{n+1} - x_n &=& \Delta + \epsilon f_z + {1\over2} \epsilon^2 f_{zz} + \cdots,
\end{eqnarray}
where subscripts again denote partial derivatives.  As we see, the function
\begin{equation}
  g(z,T) \equiv f_z(z,T)
   \label{def:g}
\end{equation}
then gives the leading-order deviation in both headway and velocity from the exact uniform-flow steady state; we will have occasion to express many of our results in terms of it.

To derive an evolution equation for $f$, we substitute the ansatz (\ref{ansatz})-(\ref{def:z,T}) into the equation of motion (\ref{startpt}) and expand in powers of $\epsilon$.  The calculation is lengthy but straightforward; carrying it out to fourth order yields (after canceling an overall factor $\epsilon^2$)
\begin{eqnarray}
 f_T = D f_{zz} &+& {1\over2} V_s'' f_z^2
  + \epsilon [C_{11T} f_{zT} + C_{11} f_{zzz} + C_{12T} f_z f_T + C_{12} f_z f_{zz} + C_{13} f_z^3] \nonumber\\*
  &+& \epsilon^2 [C_{21} f_{zzzz} + C_{22a} f_z f_{zzz} + C_{22b} f_{zz}^2 + C_{23} f_z^2 f_{zz} + C_{24} f_z^4 + \cdots]
 \label{expansion}
\end{eqnarray}
The omitted terms in the $\epsilon^2$ correction all involve $T$-derivatives of $f$, and turn out not to affect any of our analysis.  The various coefficients are given by
\begin{subequations}
\begin{eqnarray}
 D &=& (\Omega_c - V_s') \tau V_s', \label{def:D} \\
 C_{11T} &=& \lambda - 2 \tau V_s', \\
 C_{11} &=& {1+3\lambda \over 6} V_s', \\
 C_{12T} &=& \tau (A_{hv} + A_{vv} V_s') = {\tau' \over \tau}, \\
 C_{12} &=& {\tau\over2} [A_{hh} + (A_{hv}+2A_{h\dot h})V_s' + 2A_{\dot h v}V_s'^2]
  = {1\over2} (V_s'' + 2\tau V_s' \Omega_c'), \\
 C_{13} &=& {\tau\over6} (A_{hhh} + 3A_{hhv}V_s' + 3A_{hvv}V_s'^2 + A_{vvv}V_s'^3)
  = {1\over6} \left( V_s''' - {3\tau' \over \tau} V_s'' \right), \\
 C_{21} &=& {1+4\lambda \over 24} V_s', \\
 C_{22a} &=& {\tau\over6} [A_{hh} + (A_{hv}+3A_{h\dot h})V_s' + 3A_{\dot h v}V_s'^2]
  = {1\over6} \left[V_s'' + \tau V_s' \left({1+3\lambda \over \tau}\right)' \right], \\
 C_{22b} &=& {\tau\over8} (A_{hh} + 4A_{h\dot h}V_s' + 4A_{\dot h \dot h}V_s'^2) \nonumber \\*
  &=& {1\over8} [V_s'' + 4\tau V_s' \Omega_c' + \tau V_s'^2 (A_{vv}-4A_{\dot h v}+4A_{\dot h \dot h})], \\
 C_{23} &=& {\tau\over4} [A_{hhh} + 2(A_{hhv}+A_{hh\dot h})V_s' + (A_{hvv}+4A_{h\dot h v})V_s'^2 + 2A_{\dot h vv}V_s'^3] \nonumber \\*
  &=& {1\over4} \left[ V_s''' - {3\tau' \over \tau} V_s'' + 2\tau V_s' \Omega_c'' + \tau V_s' V_s'' (A_{vv}-2A_{\dot h v}) \right], \\
 C_{24} &=& {\tau\over24} (A_{hhhh} + 4A_{hhhv}V_s' + 6A_{hhvv}V_s'^2 + 4A_{hvvv}V_s'^3 + A_{vvvv}V_s'^4) \nonumber \\*
  &=& {1\over24} \left[V_s'''' - {4\tau' \over \tau} V_s''' + 6\tau\left({1\over\tau}\right)'' V_s'' + 3\tau A_{vv} V_s''^2 \right],
\end{eqnarray}
\end{subequations}
where all derivatives of $A$ are evaluated in the steady-flow state, and derivatives of $V_s$, $\tau$, $\lambda$, and $\Omega_c = (1+2\lambda)/2\tau$ are evaluated at $\Delta$.  In the OVM and FVDM, the acceleration function $A(h,\dot h,v)$ is taken to be linear in the velocities $\dot h$ and $v$, with coefficients that are independent of $h$.  For these models, then, $\tau$, $\lambda$, and $\Omega_c$ are constants, and many of the coefficients above simplify considerably.  As we will soon see, the fact that $C_{12}$ then reduces to $V_s''/2$ will prove particularly consequential.  On the other hand, features that arise due to nonzero $\Delta$-derivatives of $\tau$, $\lambda$, and $\Omega_c$ will be novel.

According to (\ref{expansion}), the position deviation $f$ obeys a Burgers equation with corrections,
\begin{equation}
  f_T = D f_{zz} + {1\over2} V_s'' f_z^2 + O(\epsilon),
\end{equation}
or, after differentiating with respect to $z$ and using the definition (\ref{def:g}),
\begin{equation}
  g_T = D g_{zz} + V_s'' g g_z + O(\epsilon),
\end{equation}
This result has been obtained previously for the OVM \cite{Nagatani-PRE2000}, the FVDM \cite{OuDaiDong-JPA2006}, and the COVM \cite{TianJiaLi-CTP2011}; the present calculation shows that the derivation of the Burgers equation, as well as the identification of the coefficients in terms of the steady velocity function $V_s$, continue to be valid even at the level of generality of (\ref{startpt}).

Near the onset of instability of steady traffic, $D$ is small.  Thus, in this regime, the neglected order-$\epsilon$ corrections might have effects comparable in size to those of the linear term in the Burgers equation.  We turn next to an examination of this situation.

\section{\label{sec:KdV}Near the onset of instability}

Suppose conditions are such that steady traffic flow is either slightly unstable or slightly stable, so that the parameter $D$ in (\ref{def:D}) is small.  Let us then write
\begin{equation}
  V_s'(\Delta) = \Omega_c(\Delta) + \epsilon^2 \delta,
\end{equation}
where $\epsilon$ is small and $\delta$ is arbitrary; we could (but will not) choose $\delta$ to be $+1$ when steady flow is slightly unstable or $-1$ when it is slightly stable, which would then give a specific meaning to the previously arbitrary small parameter $\epsilon$.  From the small-$k$ linear stability result (\ref{sigmaatonset}), we see that the unstable perturbations (for positive $\delta$) have wave numbers of order $\epsilon$.  Then, once the $i k V_s'$ in the imaginary part of the linear growth rate has been absorbed into the choice of reference frame in (\ref{def:z,T}), the linear growth rate is then of order $\epsilon^3$.  Thus we retain the definition of $z$ in (\ref{def:z,T}), but replace the definition of $T$ by
\begin{equation}
  T \equiv \epsilon^3 t.
   \label{def:KdVT}
\end{equation}
Since $V_s'$ is near the onset of instability, we find from (\ref{def:D}) that $D$ is also of order $\epsilon^2$.  Thus in the Burgers equation (\ref{expansion}), either $f$ must grow without bound or the $f_z^2$ term must eventually be balanced by the correction terms.  In the latter case, the size of $f$ must be only of order $\epsilon$, so that the quadratic term in the Burgers equation can be balanced by the linear term in the correction.  (Other correction terms, being higher powers of $f$, would be of higher order.)  Thus we modify the original ansatz (\ref{ansatz}) to read
\begin{equation}
  x_n = n\Delta + V_s(\Delta)t + \epsilon f(z,T),
\end{equation}

We may now substitute this new ansatz into the evolution equation (\ref{startpt}) and expand to fifth order in $\epsilon$.  Equivalently, however, we can simply note that the new scalings have the effect of multiplying each $f$ in (\ref{expansion}) by an additional factor of $\epsilon$ and also multiplying each $T$-derivative by an additional $\epsilon$.  After canceling an overall factor of $\epsilon^2$, this then gives, to order $\epsilon$,
\begin{equation}
  f_T = C_{11} f_{zzz} + {1\over2} V_s'' f_z^2
  + \epsilon \left[-{1+2\lambda \over 2} \delta f_{zz} + C_{11T} f_{zT} + C_{12} f_z f_{zz} + C_{21} f_{zzzz} \right].
\end{equation}
The leading order of this evolution equation is related to the Korteweg-deVries equation; if we take its z-derivative and rewrite it in terms of $g \equiv f_z$, then the leading terms take the standard KdV form
\begin{equation}
  g_T = C_{11} g_{zzz} + V_s'' g g_z
  + \epsilon \left[-{1+2\lambda \over 2} \delta g_z + C_{11T} g_T + C_{12} g g_z + C_{21} g_{zzz}\right]_z.
\end{equation}
The correction terms can be simplified somewhat.  Since $V_s'$ is close to $\Omega_c$, we easily find $C_{11T}=-(1+\lambda)$ to leading order.  Replacing the $g_T$ in the correction with the leading-order terms, substituting the explicit expressions for the coefficients, and simplifying gives
\begin{equation}
  g_T = {1+3\lambda \over 6} V_s' g_{zzz} + V_s'' g g_z
  - {1+2\lambda \over 8} \epsilon [4 \delta g + (1+2\lambda) V_s' g_{zz} + 2 (V_s'' - \Omega_c') g^2]_{zz}.
   \label{fullKdV}
\end{equation}

One-soliton solutions of the KdV equation -- the leading order of (\ref{fullKdV}) -- describe local increases or decreases in the car density that preserve their form while propagating along the line of cars.  There is a one-parameter family of these solutions, given explicitly by
\begin{equation}
  g^{(0)}(z,T;k) = M k^2 \text{sech}^2 (k z + k^3 u T),
   \label{1soliton}
\end{equation}
with $k$ being a free parameter, and
\begin{equation}
  M = 2 [1+3\lambda(\Delta)] {V_s'(\Delta) \over V_s''(\Delta)}, \qquad u = {2\over3} [1+3\lambda(\Delta)] V_s'(\Delta).
\end{equation}
Recall that $z$ specifies the location of the soliton in a reference frame moving upstream at a rate (cars per unit time) of $V_s'$, so the effect of the $u$ in this expression is to increase that rate by a factor $1+2(1+3\lambda)\epsilon^2 k^2 /3$.  The amplitude $M$ is positive if $V_s''$ is positive; in this case the soliton represents a local rarefaction of traffic, with both the headway and speed being higher in the center of the sech$^2$ lump.  On the other hand, if $V_s''$ is negative, then the soliton describes a local jam, a region where cars move more slowly and are closer together than average.  For larger $k$ the disturbance is larger in amplitude but smaller in spatial extent, and moves more rapidly through the line of traffic.  The arbitrariness of $k$ reflects the invariance of the KdV equation under the simultaneous rescalings $z \to k z$, $T \to k^3 T$, $g \to k^2 g$.  This invariance, in turn, reflects the arbitrariness of the small parameter $\epsilon$:  replacing $\epsilon$ with $k \epsilon$ implements the rescalings.

The correction term $\epsilon g_{zz} \delta$ in (\ref{fullKdV}) breaks the invariance of the equation under rescaling, and so one might expect that the corrections may break the continuous family of solutions (\ref{1soliton}) down to a discrete set, or possibly a single solution.  That is, we expect that only a discrete subset of the family of $\epsilon = 0$ solutions actually represents the $\epsilon \to 0$ limits of solutions of the full equation (\ref{fullKdV}).  Typically, one derives a solvability condition by linearizing (\ref{fullKdV}) about the zero-order solution, and the surviving solutions have $k$ values that satisfy that solvability condition.  However, it is more instructive to calculate how the correction terms cause the value of $k$ to evolve in time \cite{ZhouLiuLuo-JPA2002,KurtzeHong-PRE1995}.  The solvability condition gives no indication of whether the ``selected'' $k$ is stable or unstable, while the multiple-time-scales approach does have something to say about this.

To begin the calculation, we assume that we can write the first-order solution to (\ref{fullKdV}) in the form
\begin{equation}
  g(z,T) = g^{(0)}(z,T;k(\epsilon T)) + \epsilon g^{(1)}(z,T).
\end{equation}
That is, we allow the parameter $k$ to evolve slowly in time.  Substituting this into (\ref{fullKdV}) and expanding to first order gives
\begin{equation}
  {\cal L}g^{(1)} + {\partial g^{(0)} \over \partial k} \dot k = - {1+2\lambda \over 8} [4 \delta g^{(0)}
      + (1+2\lambda) V_s' g^{(0)}_{zz} + 2 (V_s''-\Omega_c') g^{(0)2}]_{zz},
   \label{1solitonpert}
\end{equation}
where $\dot k$ denotes the derivative of $k$ with respect to the slow time variable $\epsilon T$, and the linear operator ${\cal L}$ is given by
\begin{equation}
  {\cal L} = {\partial \over \partial T} - {1+3\lambda \over 6} V_s' {\partial^3 \over \partial z^3}
    - V_s'' {\partial \over \partial z} g^{(0)}.
\end{equation}
The fact that there is a continuous family of solutions $g^{(0)}$ of the zero-order problem suggests that there will be a function that is annihilated by the adjoint operator ${\cal L}^\dagger$, and in fact this function turns out to be $g^{(0)}$ itself.  Explicitly, with an inner product $(\cdot,\cdot)$ that consists of integrating over all $z$ and averaging over all $T$, the adjoint operator is
\begin{equation}
  {\cal L}^\dagger = -{\partial \over \partial T} + {1+3\lambda \over 6} V_s' {\partial^3 \over \partial z^3}
    + g^{(0)} V_s'' {\partial \over \partial z}.
\end{equation}
Applying this to $g^{(0)}$ gives zero because $g^{(0)}$ satisfies (\ref{fullKdV}) to leading order.  Thus if we take the inner product of $g^{(0)}$ with both sides of (\ref{1solitonpert}), the term involving $g^{(1)}$ vanishes and we are left with
\begin{equation}
   \left( g^{(0)}, {\partial g^{(0)} \over \partial k} \right) \dot k = -{1+2\lambda \over 8} [4 (g^{(0)}, g^{(0)}_{zz}) \delta
   + (1+2\lambda) V_s' (g^{(0)},g^{(0)}_{zzzz}) + 2 (V_s''-\Omega_c') (g^{(0)},(g^{(0)2})_{zz})].
\end{equation}
The usual solvability condition is simply the right hand side of this set equal to zero; it locates the $k$ values which are not moved by the correction terms.  Keeping the $\dot k$ term on the left reveals whether $k$'s near those values are driven toward or away from them.  Evaluating the integrals yields
\begin{equation}
  \dot k = {4[1+2\lambda(\Delta)]\over15} k^3 \delta
  + {4[1+2\lambda(\Delta)]\over105} V_s' \left\{ -5 [1+2\lambda(\Delta)]
   + 8 [1+3\lambda(\Delta)] {V_s''(\Delta) - \Omega_c'(\Delta) \over V_s''(\Delta)} \right\} k^5.
   \label{kevolution}
\end{equation}
If we set $\Omega_c'=0$ in this equation, then the $k$ for which $\dot k$ vanishes agrees with the solvability result of Ou et al. \cite{OuDaiDong-JPA2006} for the FVDM.

Note that $V_s''$ drops out of (\ref{kevolution}) completely if $\Omega_c'$ vanishes.  This is because for $\Omega_c'=0$, both the quadratic terms in (\ref{fullKdV}), and none of the linear terms, carry factors of $V_s''$.  As a result, $V_s''$ can be scaled out of the equation completely.  For nonzero $\Omega_c'$, on the other hand, there is a role for $V_s''$ to play.

In the OVM, the FVDM, and the COVM, $\Omega_c$ is a constant, so $\Omega_c'=0$ and the coefficient of $k^5$ on the right side of (\ref{kevolution}) is manifestly positive. Thus for $\delta>0$ -- the range for which steady traffic flow is linearly unstable -- (\ref{kevolution}) drives the value of $k$ higher and higher, in fact diverging in finite time.  This means that the actual pattern of traffic flow evolves out of the regime in which the KdV equation (\ref{fullKdV}) is valid.  On the other hand, for $\delta<0$ there is a nontrivial fixed point of (\ref{kevolution}) with $k \propto |\delta|^{1/2}$.  If $k$ starts below this, then it decreases, decaying to zero as $T^{-1/2}$, while if $k$ starts above the fixed point, then it is again driven to infinity.  Thus the ``selected'' $k$ actually marks the threshold of a {\it finite-amplitude\/} instability of steady traffic, even when steady flow is {\it linearly\/} stable.

New possibilities arise when $\Omega_c'$ is nonzero, because the coefficient of $k^5$ in (\ref{kevolution}) can be negative.  Should that be the case, there would be a nontrivial fixed point when $\delta$ is positive.  Thus when steady traffic flow is (slightly) linearly unstable, the instability leads to small-amplitude jams described by soliton solutions of the KdV equation (\ref{fullKdV}).  When steady flow is linearly stable, so $\delta<0$, then (\ref{kevolution}) would have $k$ decay to zero no matter where it started; there would be no hint of a finite-amplitude instability in this case.

Note that at the threshold of {\it absolute\/} stability we have $V_s''=\Omega_c'$.  In this case the coefficient of $k^5$ in (\ref{kevolution}) is negative.  In general, the calculation above then predicts that traffic will be in the regime where small-amplitude jams are seen, given by (\ref{1soliton}) with $k=(7\delta/10\tau V_s'^2)^{1/2}$.  This fails, however, if the threshold of absolute stability is at or near an inflection point of $V_s$, as is the case in the OVM, the FVDM, and the COVM.

\section{\label{sec:mKdV}Near an inflection point}

The picture changes dramatically if, in addition to being near the threshold of instability, the traffic spacing $\Delta$ is near an inflection point of the steady-state velocity function $V_s$.  Suppose the inflection point occurs at $\Delta=\Delta_i$, and $V_s'(\Delta_i)$ is within order $\epsilon^2$ of the instability threshold,
\begin{equation}
  V_s'(\Delta_i) = \Omega_c(\Delta_i) + \epsilon^2 \delta_i.
\end{equation}
Let $\Delta$ itself be within order $\epsilon^2$ of the inflection point, say
\begin{equation}
  \Delta = \Delta_i + \epsilon^2 \beta.
\end{equation}
It follows that $V_s''(\Delta)$ is also small, specifically
\begin{equation}
  V_s''(\Delta) = \epsilon^2 V_s'''(\Delta_i) \beta.
    \label{def:beta}
\end{equation}
Note that if $V_s'$ has a {\it maximum\/} at $\Delta_i$, then $V_s'''(\Delta)$ is negative.  Since $V_s''(\Delta_i)$ vanishes, it follows that $V_s'(\Delta)$ differs from $V_s'(\Delta_i)$ only by order $\epsilon^4$.  From (\ref{def:D}) we then find
\begin{equation}
  D = -{1+2\lambda \over 2} \epsilon^2 \delta \qquad \text{with} \qquad \delta = \delta_i - \Omega_c' \beta
    \label{def:delta}
\end{equation}
with corrections of order $\epsilon^4$.  As before, steady flow is linearly unstable for positive $\delta$, linearly stable for negative $\delta$.

From (\ref{expansion}) we see that if both $D$ and $V_s''$ are of order $\epsilon^2$, then $f_T$ is of order $\epsilon$, and the leading-order terms are the three order-$\epsilon$ terms on the right side that do not contain $T$-derivatives.  It is appropriate, then, to return to the original ansatz (\ref{ansatz}) for $x_n$, but retain the new definition (\ref{def:KdVT}) of the slow time $T$.  As in the preceding section, we may substitute all this into the evolution equation (\ref{startpt}) and expand (to fifth order in $\epsilon$), or we may simply take the general expansion (\ref{expansion}) and multiply each $T$-derivative by an additional factor $\epsilon$.  After substituting (\ref{def:delta}) for $D$ and (\ref{def:beta}) for $V_s''$ and canceling an overall factor of $\epsilon$, this latter procedure yields, to first order,
\begin{eqnarray}
 f_T = C_{11} f_{zzz} + C_{12} f_z f_{zz} + C_{13} f_z^3 + \epsilon \bigg[ &-&{1+2\lambda \over 2}f_{zz}\delta + {\beta\over2} V_s''' f_z^2 + C_{11T} f_{zT} + C_{12T} f_z f_T + C_{21} f_{zzzz} \nonumber \\*
 &+& C_{22a} f_z f_{zzz} + C_{22b} f_{zz}^2 + C_{23} f_z^2 f_{zz} + C_{24} f_z^4\bigg].
\end{eqnarray}
To simplify the corrections, we again substitute the leading-order terms for each $f_T$ in the correction.  We then differentiate with respect to $z$ and write the resulting equation in terms of $g$ to obtain
\begin{eqnarray}
 g_T = {1+3\lambda\over6} V_s' g_{zzz} &+& {1+2\lambda\over4} \Omega_c' (g^2)_{zz} + {1\over6} V_s''' (g^3)_z \nonumber \\*
  &+& \epsilon \left[-{1+2\lambda\over2} g_z\delta + {\beta\over2} V_s''' g^2 - {(1+2\lambda)^2\over8} V_s' g_{zzz} \right.  \nonumber \\*
  &~& \qquad \left. + \tilde C_{22a} g g_{zz} + \tilde C_{22b} g_z^2 + \tilde C_{23} g^2 g_z + {1\over24}V_s'''' g^4\right]_z,
 \label{fullmKdV}
\end{eqnarray}
with the new coefficients given by
\begin{subequations}
\begin{eqnarray}
  \tilde C_{22a} &=& C_{22a} + C_{12T} C_{11} + C_{11T} C_{12}
     = -{(1+2\lambda)^2\over4} \left( {1+\lambda \over \tau} \right)', \\
  \tilde C_{22b} &=& C_{22b} + C_{11T} C_{12} = {(1+2\lambda)^2\over4}
   \left[ -\Omega_c' + {1\over8\tau}(A_{vv} - 4A_{v\dot h} + 4A_{\dot h\dot h}) \right], \\
  \tilde C_{23} &=& C_{23} + 3 C_{11T} C_{13} + C_{12T} C_{12}
   = {1+2\lambda \over4} \left[ (\Omega_c - V_s')'' + 2 {\tau' \over \tau} \Omega_c' \right].
\end{eqnarray}
\end{subequations}
Note that both $\tilde C_{22a}$ and $\tilde C_{22b}$ vanish under the assumptions underlying the OVM and FVDM.

The leading order of (\ref{fullmKdV}) is the mKdV equation {\it only\/} if $\Omega_c'$ vanishes or is at most of order $\epsilon$.  Thus, the mKdV equation only appears when {\it both\/} $V_s''$ and $\Omega_c'$ are small for a given $\Delta$.  That is, the mKdV equation is {\it not\/} a generic feature of car-following models.  Rather, it appears only in those cases for which the threshold of absolute stability occurs at or near an inflection point of the steady-state velocity function.  If the inflection point and the threshold of absolute stability are not close together, then we obtain a KdV equation near the former (as we found in the preceding section) and the more general equation (\ref{fullmKdV}) near the latter.  Equation (\ref{fullmKdV}) does, however, share some important features of the mKdV equation.

In those cases where the mKdV equation does apply, its hyperbolic-tangent ``kink'' solutions are of particular interest.  In the context of traffic, a one-kink solution describes a ``domain wall'' separating regions of uniform traffic flow with different speeds and densities.  The mKdV equation has a one-parameter continuous family of such solutions, and one can carry out a solvability calculation to determine which are preserved by the correction terms.  Like the mKdV equation, the full equation (\ref{fullmKdV}) also admits a one-parameter family of kink solutions.  Remarkably, in the general case it is also possible to carry out the perturbation calculation to find members of the family that persist when the order-$\epsilon$ corrections are included.

One-kink solutions of the leading order of (\ref{fullmKdV}) are given by
\begin{equation}
  g^{(0)}(z,T;k) = M k \tanh (k z + k^3 u T),
    \label{1kink}
\end{equation}
where $M$ and $u$ are given by
\begin{equation}
  V_s''' M^2 - 3 (1+2\lambda) \Omega_c' M + 2 (1+3\lambda) V_s' = 0, \qquad u = {1\over6} V_s''' M^2.
\end{equation}
Since $V_s'''$ is negative, there are two solutions for $M$, one positive and one negative, and -- $u$ being negative -- both move relative to traffic slightly more slowly than $V_s'$.  As was the case for the KdV solitons, the various powers of $k$ in (\ref{1kink}) reflect the invariance of the leading-order evolution equation under the simultaneous rescalings $z \to k z$, $T \to k^3 T$, $g \to k^{-1} g$, which in turn reflects the arbitrariness of the small parameter $\epsilon$.

To see the effect of the order-$\epsilon$ corrections on the kink solutions, we would like to carry out a perturbation calculation analogous to the one in the preceding section.  That is, we write the first-order solution as
\begin{equation}
  g(z,T) = g^{(0)}(z,T;k(\epsilon T)) + \epsilon g^{(1)}(z,T),
\end{equation}
allowing the parameter $k$ to evolve slowly in time, and substitute into (\ref{fullmKdV}) to get
\begin{eqnarray}
   {\cal L}g^{(1)} + {\partial g^{(0)} \over \partial k} \dot k  = 
   \bigg[&-&{1+2\lambda\over2}g^{(0)}_z\delta + {\beta\over2} V_s''' g^{(0)2} - {(1+2\lambda)^2\over8} V_s' g^{(0)}_{zzz} \nonumber \\*
    &+&\tilde C_{22a}g^{(0)}g^{(0)}_{zz}+\tilde C_{22b}(g^{(0)}_z)^2+\tilde C_{23}g^{(0)2}g^{(0)}_z+{1\over24}V_s''''g^{(0)4}\bigg]_z.
     \label{1kinkpert}
\end{eqnarray}
The linear operator ${\cal L}$ is now
\begin{equation}
  {\cal L} = {\partial \over \partial T} - {1+3\lambda \over 6} V_s' {\partial^3 \over \partial z^3}
    - {(1+2\lambda) \over 2} \Omega_c' {\partial^2 \over \partial z^2} g^{(0)}
    - {1\over2} V_s''' {\partial \over \partial z} g^{(0)2}.
\end{equation}
Next we seek the function $s(z,T)$ which is annihilated by the adjoint operator ${\cal L}^\dagger$,
\begin{equation}
  0 = {\cal L}^\dagger s = -s_T + {1+3\lambda \over 6} V_s' s_{zzz} - {(1+2\lambda) \over 2} \Omega_c' g^{(0)} s_{zz}
    + {1\over2} V_s''' g^{(0)2} s_z.
\end{equation}
The equation can, in fact, be solved using standard techniques; we find that $s$ is given by
\begin{equation}
  s = s(k z + k^3 u T) \qquad \text{with} \qquad {ds(y) \over dy} = \text{sech}^{2+p} y,
\end{equation}
the exponent $p$ being given by
\begin{equation}
  p = - {3 M (1+2\lambda) \Omega_c' \over (1+3\lambda) V_s'}.
\end{equation}
Note that for $\Omega_c'=0$ we have simply $p=0$, so $s$ reduces to $g^{(0)}$ as in the preceding section.

We now take the inner product of the function $s$ with both sides of (\ref{1kinkpert}) to obtain the evolution equation for the parameter $k$,
\begin{eqnarray}
   \left( s,{\partial g^{(0)} \over \partial k} \right) \dot k  = 
   \bigg(s, \bigg[&-&{1+2\lambda\over2}g^{(0)}_z\delta + {\beta\over2} V_s''' g^{(0)2} - {(1+2\lambda)^2\over8} V_s' g^{(0)}_{zzz} \nonumber \\*
   &+& \tilde C_{22a} g^{(0)} g^{(0)}_{zz} + \tilde C_{22b} (g^{(0)}_z)^2 + \tilde C_{23} g^{(0)2} g^{(0)}_z + {1\over24}V_s'''' g^{(0)4}\bigg]_z\bigg).
\end{eqnarray}
On the right side, we can change variables to $y = k z + k^3 u T$, integrate by parts (noticing that the boundary term does not vanish, because neither $s$ nor the $g^{(0)2}$ and $g^{(0)4}$ terms in the square bracket vanish at infinity), and evaluate the resulting integrals explicitly in terms of gamma functions.  Unfortunately the integral on the left side of the equation diverges, whether or not $\Omega_c'$ vanishes.  Setting the right side of the equation to zero -- the solvability condition -- thus identifies the single $k$ value for which the $g^{(0)}$ given by (\ref{1kink}) is the $\epsilon \to 0$ limit of the perturbed solution.  Specifically, this yields
\begin{eqnarray}
  (1+2\lambda) \delta + V_s''' M \beta = {k^2 \over 5+2p} \bigg\{ (1+2\lambda)^2 (1+p) V_s' &+& 4 \bigg[ (2+p) \tilde C_{22b} - \tilde C_{22a} \bigg] M + 2 \tilde C_{23} M^2 \nonumber \\*
  &-& {1\over6} (3+p) V_s'''' M^3 \bigg\}.
\end{eqnarray}
For the FVDM, $p$, $\tilde C_{22a}$, and $\tilde C_{22b}$ all vanish, and $\tilde C_{23}$ is equal to $-(1+2\lambda)V_s'''/4$.  The result above then agrees with the result given by Ou et al. \cite{OuDaiDong-JPA2006}, provided we set $\beta$ and $V_s''''$ to zero, as they do (explicitly for $\beta$ and implicitly for $V_s''''$).

For $k$ values which do not satisfy the solvability equation, the above perturbation calculation is unable to determine what effect the correction terms have on the solution.  Thus the solvability approach is silent on the question of how, and indeed whether, the ``selected'' kink is established.

\section{\label{sec:disc}Discussion}

We have examined a very general car-following model embodied in Equation (\ref{startpt}), in which each car's acceleration is a general function of its current speed, the headway between it and the next car ahead of it, and the rate of change of the headway, subject to monotonicity assumptions that rule out unreasonable descriptions of driver behavior and an assumption that sufficiently many derivatives exist.  The most popular previously published models in this class, those for which the more delicate nonlinear calculations described below have been carried out, have assumed that the dependence of acceleration on the velocities is {\it linear}, with coefficients that are {\it independent\/} of headway \cite{Bando-PRE1995,JiangWuZhu-PRE2001}.

Most car-following models are based on an ``optimal velocity function'' $V_s(\Delta)$, written explicitly into the model, which gives the speed of uniform steady-state traffic flow in terms of the headway $\Delta$ between cars.  In those models, uniform steady flow is found to be linearly unstable if the steady-state traffic speed is too sensitive to headway, i.e. if the derivative $V_s'(\Delta)$ exceeds some limit $\Omega_c$ which can be calculated in terms of the parameters of the model.  It is also possible to have ``absolute stability,'' which is the situation where $V_s'(\Delta)$ remains below the instability limit for all $\Delta$.  The threshold of absolute stability is reached when parameter values are such that the {\it maximum\/} value of $V_s'(\Delta)$ matches the instability limit $\Omega_c$; in the most popular models $\Omega_c$ is a constant, so this can only occur at an inflection point of the optimal velocity function.  In the more general setting, the optimal velocity function must be defined implicitly from the acceleration function, as in (\ref{def:Vs}).  The linear stability results for uniform, steady flow then continue to hold in the general case, with the crucial difference that the instability limit $\Omega_c$ can itself be a nontrivial function of the headway $\Delta$.  This implies, in particular, that the threshold of absolute stability does {\it not\/} generally coincide with an inflection point of the optimal velocity function.

For steady traffic with a spacing near the onset of instability, linear stability analysis reveals that the perturbations that are unstable (or least stable) have long wavelengths.  Using this observation, one often reduces a car-following model in this regime to a continuum model, in which the spacing between cars is governed by a Burgers equation plus small corrections.  In more restricted parameter ranges, different evolution equations for the traffic flow result:  the Korteweg-deVries (KdV) equation (plus corrections) near the onset of instability, the modified Korteweg-deVries (mKdV) equation (plus corrections) near the threshold of absolute stability.  The KdV equation admits a one-parameter family of soliton solutions, and the mKdV equation a one-parameter family of kink solutions; conventionally one carries out solvability analyses that determine unique soliton and kink solutions from among these families that are preserved in the presence of the correction terms.  This program of reduction to nonlinear evolution equations has been carried out for a number of car-following models \cite{TianJiaLi-CTP2011,Nagatani-PRE2000,OuDaiDong-JPA2006,MuramutsuNagatani-PRE1999,GeChengDai-PhysA2005,KomatsuSasa-PRE1995} that are special cases of (\ref{startpt}), as well as for lattice models \cite{Nagatani-PRE1999,YuChengGe-CPB2010} and models which incorporate finite time delays \cite{GeChengDai-PhysA2005,Nagatani-PRE1998,NagataniNakanishiEmmerich-JPA1998,NagataniNakanishi-PRE1998}, look-ahead \cite{ShiChenXue-CTP2007,TangHuangWongJiang-ActaMechSin2008}, look-back \cite{HayakawaNakanishi-PRE1998}, and a dependence on the acceleration of the next car \cite{TianJiaLiGao-CPB2010}.  All in all, it is not surprising that the derivations should continue to hold with those extra effects included, since the starting point of each derivation is the assumption that the traffic flow evolves on length scales long compared to the headway $\Delta$ and time scales long compared with the time scale on which an individual driver adjusts speed.

We find that the derivation of the Burgers equation goes through at the level of generality of (\ref{startpt}), and that the coefficients in the resulting equation match those that have been obtained in the more specific models in the past.  In particular, the coefficient of the nonlinear term in the Burgers equation is simply given by $V_s''(\Delta)$.

Near the onset of instability of steady flow, we find that the derivation of the Korteweg-deVries equation also goes through as in the more specific models, but if the instability limit $\Omega_c$ is a nontrivial function of headway $\Delta$, then there is a crucial change to the coefficient of one of the correction terms which can lead to qualitative changes in the nature of the solutions.  Specifically, if $\Omega_c'(\Delta)$ vanishes, as in the previous models, then a solvability analysis of the one-soliton solutions of the KdV equation shows that there is a ``selected'' soliton only when uniform, steady flow is linearly stable.  However, when that derivative is nonzero -- and close enough to $V_s''(\Delta)$ -- then the opposite may be the case: a selected soliton exists only when the uniform flow is linearly {\it unstable}.

Rather than using the usual solvability condition, we prefer to use a multiple-time-scales approach \cite{KurtzeHong-PRE1995,ZhouLiuLuo-JPA2002} to investigate the effect of the correction terms on the family of one-soliton solutions of the KdV equation.  The advantage of this approach is that it not only identifies the ``selected'' soliton, but also shows what happens to the other, non-selected members of the family.  Namely, the corrections drive a slow evolution of the soliton parameter, at a rate that vanishes for the selected soliton.  The analysis shows that there are two possible behaviors.  In one scenario, there is a selected soliton only when steady flow is linearly stable, and this selected soliton is unstable: smaller solitons decay to zero, larger ones grow.  In this scenario, which is the one that obtains when $\Omega_c'(\Delta)$ vanishes, the selected soliton marks the threshold of a {\it finite-amplitude\/} instability of the {\it linearly\/} stable steady traffic flow.  Beyond this threshold amplitude, which goes to zero as parameters are adjusted toward the threshold of linear instability, jams grow large enough to invalidate the assumptions underlying the derivation of the KdV equation.  In the other scenario, a selected soliton exists only when steady traffic flow is linearly unstable, and the slow evolution of the soliton parameter that is caused by the correction terms drives the system toward this selected soliton.  That is, there is a forward bifurcation from stable steady flow to a stable, small-amplitude jam.  In either scenario, the evolution of the soliton is slow.  Specifically, if the derivative of the optimal velocity function differs from its critical value by a small amount $\epsilon$, then the soliton adjustment takes place on a time scale of $\epsilon^{-2}$.  Thus, even if the soliton is in the process of decaying to zero, it can {\it appear\/} to be a persistent feature of the traffic flow.

When parameter values are such that steady flow is near the onset of instability {\it and\/} near an inflection point of the optimal velocity function, then the assumptions underlying the derivation of the KdV equation are not satisfied.  In previous models this is what occurs near the threshold of absolute stability, and in this regime one reduces the model to a modified Korteweg-deVries equation with corrections \cite{KomatsuSasa-PRE1995,OuDaiDong-JPA2006}.  We find, however, that this derivation does not go through in the general case.  Rather, the result of the derivation is the mKdV equation plus an extra term.  The existence of this term had been anticipated by Komatsu and Sasa \cite{KomatsuSasa-PRE1995}, and it had been seen by Hayakawa and Nakanishi \cite{HayakawaNakanishi-PRE1998} in a model including look-back; in our case the coefficient of that extra term is proportional to $\Omega_c'(\Delta)$.  Thus the mKdV equation only arises if this derivative is zero, or at least small, at the inflection point of $V_s$ -- that is, if the threshold of absolute stability occurs near the inflection point.  In the previous models, $\Omega_c$ was a constant, so this happened automatically; as we pointed out above, it is not the case generically.

Like the mKdV equation, the new equation also admits a one-parameter family of kink solutions.  Moreover, it is still possible to carry out a solvability analysis for the new equation to determine which of these kink solutions are preserved when the correction terms are included.  The multiple-time-scales approach, however, fails -- as it does for the mKdV equation -- because the slow time derivative of the kink parameter is multiplied by a coefficient that turns out to diverge.  This divergence, in turn, seems to come from the fact that changing the kink parameter involves changing the asymptotic values of the kink, i.e. changing the traffic spacing infinitely far from the location of the kink itself.  This, and other aspects of the mKdV equation, will be the focus of future work.

As noted above, the threshold of absolute stability occurs where the derivative of the optimal velocity function, $V_s'(\Delta)$, matches the instability limit, $\Omega_c$, at only one value of the headway $\Delta$, and is below $\Omega_c$ for all other headway values.  We find that the behavior of traffic near the threshold of absolute stability can then be described by one of two reduced equations.  If the threshold occurs near an inflection point of the optimal velocity function, that is if both $V_s''(\Delta)$ and $\Omega_c'(\Delta)$ are small, then the model can be reduced to a mKdV equation.  This is the situation which has been explored in previous models.  If, on the other hand, the threshold occurs {\it away\/} from an inflection point, then $V_s''(\Delta)$ and $\Omega_c'(\Delta)$ are close together, but neither is small.  In this situation the KdV equation applies, and this is the regime in which the correction terms lead to a selected, small-amplitude soliton, representing a localized, small-amplitude ``jam'', when steady traffic flow is linearly unstable.

In conclusion, we wish to point out that although the {\it forms\/} of the terms appearing in reduced equations like the KdV and mKdV are determined only by the scalings that go into their derivation, the {\it coefficients\/} of those terms are sensitive to details of the model.  The behavior of the solutions of those reduced equations can be sensitive to the values, and especially to the signs, of those coefficients.  Thus {\it implicit\/} assumptions made when formulating the model can constrain the coefficients (often constraining them to be zero!), and this can then have important quantitative and qualitative effects on the kind of behavior predicted by the resulting reduced equations.

\end{document}